\documentclass[12pt]{article}
\begin{document}

\sf

\begin{center}
   \vskip 2em
  {\LARGE \sf Chern-Simons theory
  and three-dimensional surfaces}
\vskip 3em
{\large \sf  Jemal Guven \\[2em]}

\em{
Instituto de Ciencias Nucleares\\
 Universidad Nacional Aut\'onoma de M\'exico\\
Apdo. Postal 70-543, 04510 M\'exico, DF, MEXICO}
\end{center}

 \vskip 1em

\begin{abstract}
\sf There are two natural Chern-Simons theories associated with the
embedding of a three-dimensional surface in Euclidean space; one is
constructed using the  induced metric connection -- it involves only
the intrinsic geometry, the other is extrinsic and uses the
connection associated with the gauging of normal rotations. As such,
the two theories appear to describe very different aspects of the
surface geometry. Remarkably, at a classical level, they are
equivalent. In particular, it will be shown that their stress
tensors differ only by a null contribution.
Their Euler-Lagrange equations provide
identical constraints on the normal curvature. A new identity for
the Cotton tensor is associated with the triviality of the
Chern-Simons theory for embedded hypersurfaces implied by this
equivalence.
\end{abstract}


\vskip 1em


\vskip 3em

\section{\Large \sf Introduction}

Consider a three-dimensional surface in an ambient Euclidean (or
Minkowski) space. In the spirit of Deser, Jackiw and Templeton's
work on three dimensional gravity in the early eighties, one may
associate a Chern-Simons theory with the intrinsic geometry of this
surface \cite{JackiwDeser}. In contrast to their model, where the
dynamical variables are provided by the metric, in a theory of
surfaces the relevant metric is the one induced by the embedding of
the surface in space. Thus the appropriate independent dynamical
variables are the functions which describe this embedding. This
difference is reflected in the Euler-Lagrange equations: whereas the
Euler-Lagrange derivative with respect to the metric is given by the
Cotton tensor \cite{Cotton,ADMYork}, its counterpart for an embedded
surface is a contraction of this tensor with the extrinsic
curvatures as given by (\ref{eq:CK}).

\vskip1pc
In general, however, the content of these equations depends
sensitively on the number of extra dimensions. In particular, if the
surface is embedded as a hypersurface, one can show that the single
contraction vanishes identically. There are thus no local degrees of
freedom associated with this limit of the theory; the Chern-Simons
theory induced by this embedding, unlike the theory based on a
metric on which it is modeled, is trivial as a physical theory.

\vskip1pc The Cotton tensor is associated with the intrinsic
geometry of the surface. The vanishing of its contraction with the
extrinsic curvature on a hypersurface is a consequence of the fact
that the intrinsic geometry is constrained by its spatial
environment. This property of the Cotton tensor is captured by the
statement that the stress associated with a Chern-Simons theory of
hypersurfaces is null, neither transmitting physical forces nor
carrying momentum. As such, it may be constructed as the curl of a
potential. This potential will be constructed explicitly for the
hypersurface limit of the Chern-Simons theory.

\vskip1pc If the number of extra dimensions is increased the physics
described by the Chern-Simons action is not trivial. It is possible
to show that local degrees of freedom are associated with the
twisting of the surface in a five or higher-dimensional space. This
is characterized by the extrinsic twist or normal curvature
associated with gauging the rotations of the normal vectors; it thus
vanishes on a hypersurface. The Euler-Lagrange equations may be
rewritten to reflect the fact that they provide a constraint on this
curvature.

\vskip1pc The normal curvature itself is a source of interesting
geometrical invariants.  In a recent paper, properties of the
obvious Yang-Mills functional were described \cite{twist}. There
are, however, other invariants that are peculiar to particular
dimensions: one such invariant, characterizing two-dimensional
surfaces embedded in four-dimensions, is the integral of the normal
curvature itself; it was introduced by Polyakov in the context of a
limit of QCD described by strings \cite{Polyakov}.

\vskip1pc It is also evident that, for three dimensional surfaces,
there is a Chern-Simons theory associated with the normal
connection. This is a theory that appears to describe an aspect of
the surface geometry that is unrelated to the theory based on the
Christoffel connection. A particularly effective way to examine it
is to introduce auxiliary variables adapted to the connection
\cite{auxil} (as worked out in \cite{twist}). Remarkably, at a
classical level, the two theories turn out to be equivalent. In
particular, it will be shown that their stress tensors differ only
by a null contribution. Their Euler-Lagrange equations thus provide
identical constraints on the normal curvature of $\Sigma$. The two
sets of equations, (\ref{eq:CK}) and (\ref{eq:ELn}), are one and the
same.

\vskip1pc It is not too surprising that these two Chern-Simons
theories are related.  Even if  the normal curvature describes a
facet of the surface geometry that is different from the one
described by the extrinsic curvature, it is not independent of the
latter: the common dynamical variables are the embedding functions
of the surface. Just as the Gauss-Codazzi equations determine the
intrinsic Riemann curvature, the Ricci integrability conditions
determine the normal curvature completely in terms of a quadratic in
the extrinsic curvature \cite{Spivak}. What is unexpected is that
the two theories are identical as classical theories. The challenge
is to understand why.

\section{\Large \sf Chern-Simons theory, hypersurfaces and null stresses}

\vskip1pc Consider a three-dimensional surface (worldsheet) $\Sigma:
u^A \to {\bf X}(u^A)$ embedded in a $3+N$ -dimensional Euclidean
(Minkowski) space, ${\bf X}= (X^1,\cdots, X^{3+N})$, $N\ge 1$. Let
some action $S[{\bf X}]$ be associated with this surface. The
response of $S$ to a deformation of the surface ${\bf X}\to {\bf X}
+ \delta {\bf X}$ is characterized by a stress `tensor' ${\bf f}^A$.
The notation reflects the fact that ${\bf f}^A$ form $3+N$ surface
vector fields, one for each spatial extension. The invariance of
$S[{\bf X}]$ under ambient Euclidean motions (Lorentz
transformations) implies that \cite{ACG}
\begin{equation}
\delta S = - \int dV\, {\bf f}^A \cdot \partial_A \delta {\bf X}\,,
\label{eq:delS}
\end{equation}
where $dV$ is the volume element induced on $\Sigma$.  The dot
represents the usual Euclidean (Lorentz) invariant inner product on
the background. A momentum density, given by $l_A {\bf f}^A$, may be
associated with any unit (timelike) vector field, $l_A$, on
$\Sigma$. On classical membrane trajectories, ${\bf f}^A$ is
covariantly conserved:
\begin{equation}
\nabla_A {\bf f}^A=0 \,, \label{eq:conserv}
\end{equation}
 where
$\nabla_A$ is the covariant derivative compatible with the induced
metric $g_{AB}$ on $\Sigma$. ${\bf f}^A$ is the Noether current
associated with the translational invariance of $S$. There is,
however, an ambiguity inherent in the definition of ${\bf f}^A$.
This is because, within the surface, ${\bf f}^A$ is defined only
modulo an additional `null' stress ${\bf h}^A$ of the form
\begin{equation}
{\bf h}^A = \epsilon^{ABC} \partial_B {\bf B}_C\,, \label{eq:null}
\end{equation} where $\epsilon^{ABC}$ is the surface
Levi-Civita tensor.

\vskip1pc For the action $S[{\bf X}]$, consider the Chern-Simons
action associated with the intrinsic geometry of $\Sigma$,
\begin{equation}
S_{\rm CS}[A_A] =  \int dV\, \epsilon^{ABC}\,{\rm Tr}\,
\left({1\over 2}\,A_A
\partial_B A_C + {1\over 3}\, A_A A_B A_C\right)
\,, \label{eq:CS1}
\end{equation}
where $(A_A)^B{}_C = \Gamma^B_{AC}$, and the $\Gamma^B_{AC}$ are the
Christoffel connections constructed with the induced metric $g_{AB}$
on $\Sigma$ \cite{JackiwDeser}. The connection is thus itself
completely determined by ${\bf X}$: $\Gamma^C_{AB}  = \Gamma^C_{AB}
[{\bf X}]$.
The metric dependence of $S$ occurs only through $\Gamma^B_{AC}$.

\vskip1pc It is well known that if $S$ is treated as a functional of
$g_{AB}$, then
\begin{equation}
\delta S_{\rm CS}[g_{AB}] =- \int dV\, C^{AB}\delta g_{AB}\,,
\label{eq:delSCS}
\end{equation}
where the symmetric tensor
\begin{equation}
C^{AB} =  \epsilon^{ACD} \nabla_C \left({\cal R}_{D}{}^B -{1\over 4}
g_D{}^B {\cal R}\right) \label{eq:cotton}
\end{equation}
is the Cotton tensor, defined in terms of the Ricci tensor ${\cal
R}_{AB}$ and the scalar curvature ${\cal R}$ \cite{Cotton}. This
tensor is thus identified as (1/2 times) the metric stress tensor $-
(2/\sqrt{g})\, \delta S_{\rm CS}/\delta g_{AB}$. $C^{AB}$ is
conformally invariant. It vanishes if (and only if) the geometry is
flat. We are, however, interested in $S$ not as a functional of
$g_{AB}$ but as a functional of ${\bf X}$. Thus $g_{AB}= {\bf
e}_A\cdot {\bf e}_B$ is the induced metric, where the ${\bf e}_A=
\partial_A {\bf X}$ are the tangent vector fields along the
parameter curves. The deformation of the metric, in response to a
deformation ${\bf X}\to {\bf X} + \delta {\bf X}$ of the surface, is
given by $\delta g_{AB}= 2 {\bf e}_{(A}\cdot
\partial_{B)} \delta {\bf X}$. Thus, the stress associated with
$S_{\rm CS}$ is identified as
\begin{equation}
{\bf f}^A_{\rm CS}=  2 C^{AB} {\bf e}_B\,.
\end{equation}
Let the $N$ normal vectors be labeled $\{{\bf n}^1,\dots,{\bf
n}^N\}$; and suppose that ${\bf n}^I\cdot {\bf n}^J = \delta^{IJ}$.
The Gauss equations
\begin{equation}
\nabla_A {\bf e}_B =  - K_{AB}^I \,{\bf n}_I\,, \label{eq:G}
\end{equation}
define $N$ extrinsic curvature tensors $K_{AB}^I$, $I=1,\dots, N$.
These tensors are symmetric.

\vskip1pc Using Eq.(\ref{eq:G}) it is possible to expand $\nabla_A
{\bf f}_{\rm CS}^A$ as a linear combination of the tangent vectors
adapted to the surface, $\{{\bf e}_A,{\bf n}^I\}$:
\begin{equation}
\nabla_A {\bf f}_{\rm CS}^A = 2 \nabla_A C^{AB}\, {\bf e}_B  - 2
C^{AB} K_{AB}^I\, {\bf n}_I \,. \label{eq:nabf}
\end{equation}
The conservation of the Cotton tensor,
\begin{equation}
\nabla_B C^{AB}=0\,, \label{eq:Ccon}
\end{equation}
emerges as a consequence of the reparametrization invariance of
$S_{\rm CS}$. The Euler-Lagrange equations are captured by the
vanishing of the normal component of $\nabla_A {\bf f}_{\rm CS}^A$
\begin{equation}
 C^{AB} K_{AB}^I=0\,.
\label{eq:CK}
\end{equation}

\vskip1pc It is useful to place Eqs.(\ref{eq:Ccon}) and
(\ref{eq:CK}) in context. In general, if it is possible to express
an action $S[{\bf X}]$ as a functional of the induced metric and the
Riemann tensor, the conserved stress tensor can be cast as ${\bf
f}^A= T^{AB} {\bf e}_B$. The metric stress tensor $T^{AB}$ is
conserved, $\nabla_A T^{AB}=0$, as a consequence of the surface
reparametrization invariance of $S$. This is completely analogous to
the conservation of the Einstein tensor $G^{AB}$ which follows from
the reparametrization invariance of the Hilbert-Einstein action. The
equations $T^{AB} K_{AB}^I=0$, however, describe the classical
trajectories of surfaces. For example, in Regge-Teitelboim gravity,
the Hilbert-Einstein action induced by the embedding describes a
theory of four dimensional worldsheets. The vacuum Einstein
equations are replaced by the equations, $G^{AB} K^I_{AB}= 0$
\cite{ReggeT}.

\vskip1pc Let us first look at a surface embedded as a (timelike)
hypersurface with a single (spacelike) normal, ${\bf n}$ and
corresponding curvature $K_{AB}$.  Then $C^{AB} K_{AB}$ vanishes
identically in Euclidean space. ${\bf f}_{\rm CS}^A$ is thus a null
stress.

\vskip1pc This identity  involves the surface integrability
conditions in an essential way. Pulling $K_{AB}$ into the derivative
appearing in Eq.(\ref{eq:cotton}), one has
\begin{equation}
 C^{AB} K_{AB} =
  \epsilon^{ACD} \nabla_C \left[ \left({\cal R}_{D}{}^B -{1\over
4} g_D{}^B {\cal R} \right)\, K_{AB}\right] - \epsilon^{ACD}
\left({\cal R}_{D}{}^B -{1\over 4} g_D{}^B {\cal R} \right)
\nabla_C\, K_{AB}\,. \label{eq:CKiD}
\end{equation}
The Gauss-Codazzi equations
\begin{equation}
{\cal R}_{AB}= K K_{AB}- K_{AC} K^{C}{}_B \label{eq:gc0}
\end{equation}
express the Riemann (or Ricci) tensor in terms of the extrinsic
curvature. This permits a presentation of the tensor $P_{AB}= ({\cal
R}_{B}{}^C - g_B{}^C R/4)\, K_{AC}$ as a polynomial in $K_{AB}$ and
$g_{AB}$. As such, it is manifestly symmetric in $A$ and $B$.  Thus
the first term vanishes. Also the Codazzi-Mainardi equations,
\begin{equation}
\nabla_A K_{BC} - \nabla_B K_{AC}=0 \,, \label{eq:cm0}
\end{equation}
imply that $\nabla_C\, K_{AB}$ is symmetric in $A$ and $C$. Thus the
second term also vanishes. One concludes that
\begin{equation}
C^{AB} K_{AB}=0 \label{eq:CKoff}
\end{equation}
for any hypersurface.

\vskip1pc There is thus no bulk response in $S_{\rm CS}$ to a
deformation of $\Sigma$ as a hypersurface.

\vskip1pc By linking intrinsic geometry to its environment, this
identity has no analogue in Deser, Jackiw and Templeton's metric
framework. It is, however, still curious in view of its elementary
nature that it does not feature in the literature. Although
Chern-Simons theory directs us to its existence, it is worth
emphasizing that it is a purely geometrical identity, tying together
the Gauss-Codazzi and the Codazzi-Mainardi equations (\ref{eq:gc0})
and (\ref{eq:cm0}) in a subtle way. It also raises a host of
interesting questions of a geometrical nature. Eq.(\ref{eq:CKoff})
is a very strong constraint: is $C^{AB}$ the only tensor satisfying
a relation of this form? It is clear that there is no finite
polynomial in $K_{AB}$ itself that is `orthogonal' to $K_{AB}$.
Derivatives of $K_{AB}$ are required. There is good reason to
believe (see the appendix) that in three-dimensions, the Cotton
tensor is the unique symmetric tensor involving first derivatives
satisfying Eq.(\ref{eq:CKoff}). It is simple to discount the
existence of an analogous tensor for two-dimensional surfaces. It
would be interesting to determine if an analogue exists in higher
odd dimensions (where Chern-Simons theories exist). And it would be
interesting to know if (higher derivative) counterparts exist for
higher dimensional surfaces, even or odd.

\vskip1pc\noindent For completeness, in the appendix, we will show
how the null stress for a hypersurface may be reconstructed from a
potential.

\section{\Large \sf Adding co-dimensions}

\vskip1pc If $\Sigma$ is not a hypersurface, Eq.(\ref{eq:CK}) is no
longer generally valid identically. It is instructive to dismantle
the equation using surface theory to understand its content.

\vskip1pc The counterpart of the Gauss equations for the normal
vectors is provided by the Weingarten equations \cite{Spivak}
\begin{equation}
\tilde \nabla_A {\bf n}^I = K_{AB}^{I}  \,{\bf e}^B\,. \label{eq:W}
\end{equation}
For co-dimensions higher than one, these equations involves the
$SO(N)$ covariant derivative $\tilde \nabla_A$, defined by
\cite{Guven93}
\begin{equation}
\tilde\nabla_A {\bf n}^I = \nabla_A {\bf n}^I + \omega^{I}{}_J{}_{A}
{\bf n}^J\,,
\end{equation}
where the normal connection $\omega^{IJ}{}_{A}$ is given by
\begin{equation}
\omega^{IJ}{}_{A} = {\bf n}^I\cdot \partial_A {\bf n}^J =
-\omega^{JI}{}_A\,.
\end{equation}
$\omega^{IJ}{}_{A}$ may be identified with the $SO(N)$ connection
associated with the gauging of normal rotations. Its curvature,
satisfying
\begin{equation}
[\tilde \nabla_A,\tilde\nabla_B]\, {\bf n}^I =
\Omega_{AB}{}^{I}{}_{J} {\bf n}^J \,,
\end{equation}
is given by
\begin{equation}
\Omega_{AB}{}^{IJ}= \partial_A \omega^{IJ}{}_B +
\omega^I{}_K{}_A\,\omega^{KJ}{}_B - (A\leftrightarrow B)\,.
\end{equation}
\vskip1pc The intrinsic and the extrinsic  geometries of $\Sigma$
are related by integrability conditions. The generalizations of  the
Gauss-Codazzi and Codazzi-Mainardi equations for a hypersurface are:
\begin{eqnarray} {\cal R}_{ABCD} -
K_{AC\,I} K_{BD}^I + K_{AD\,I} K_{BC}^I &=& 0 \,,
\label{eq:gauss1}\\
\tilde\nabla_A K_{BC}^I - \tilde\nabla_B K_{AC}^I &=& 0\,.
\label{eq:cm1}
\end{eqnarray}
A third set of equations constrains $\Omega_{AB\,IJ}$:
\begin{equation}
\Omega_{AB\,IJ} - K_{AC\,I} K^C{}_{B\,J} + K_{AC\,J} K^C{}_{B\,I}
=0\,.\label{eq:ricci}
\end{equation}
Thus the normal curvature $\Omega_{AB}{}^{IJ}$, like the Riemann
curvature ${\cal R}_{ABCD}$, is determined completely by the
extrinsic curvature,  $K^I_{AB}$. If $K_{AB}^I$ vanishes in all but
one direction, $\Omega_{AB\,IJ}$ will also vanish.

\vskip1pc Now one has, replacing Eq.(\ref{eq:CKiD}),
\begin{equation}
 C^{AB} K_{AB}^I =
 \epsilon^{ACD} \tilde\nabla_C \left[ \left({\cal R}_{D}{}^B
-{1\over 4} g_D{}^B {\cal R} \right)\, K_{AB}^I\right] -
\epsilon^{ACD} \left({\cal R}_{D}{}^B -{1\over 4} g_D{}^B {\cal R}
\right) \tilde \nabla_C\, K_{AB}^I\,.
\end{equation}
The second term again vanishes identically because of the symmetry
implied by the Codazzi-Mainardi equations (\ref{eq:cm1}). However,
the first term does not generally vanish if $N>1$. Instead, one
possesses the identity
\begin{eqnarray}
\left({\cal R}_{D}{}^B -{1\over 4} g_D{}^B {\cal R} \right)\,
K_{AB}^I &=&
K_D{}^{B\,J} K_J K_{AB}^I  - K_{DE}^ JK^{EB}{}_J K_{AB}^I\nonumber\\
&=& {1\over 2}\left(\Omega_{AD}{}^{IJ} K_J - \Omega_{AE}{}^{IJ}
K^{E}{}_{D\,J} \right)\,,
\end{eqnarray}
where the first line follows from the Gauss-Codazzi equations
(\ref{eq:gauss1}), and the second from Eq.(\ref{eq:ricci}) on
swapping normal partners. Thus, using the Bianchi identities,
$\nabla_{[C} \Omega_{AB]}{}^{IJ} =0$, one may express
\begin{equation}
 C^{AB} K_{AB}^I =
 {1\over 2} \epsilon^{ACD}
 \left(\Omega_{AD}{}^{IJ} \tilde \nabla_C K_J - \tilde\nabla_C \Omega_{AE}{}^{IJ}
K^{E}{}_{D\,J} \right)\,, \label{eq:ckomega}
 \end{equation}
 which generally does not vanish.
The Euler-Lagrange equations Eqs.(\ref{eq:CK}) thus place
constraints on $\Omega_{AB}{}^{IJ}$. If $\Omega_{AB}{}^{IJ}$
vanishes, as it will for a hypersurface, the right hand side of
Eq.(\ref{eq:ckomega}) vanishes.

\section{\Large \sf Chern-Simons theory and the normal connection}

The embedded analogue of Deser, Jackiw and Teitelboim's action is
not the only Chern-Simons theory one can construct with the surface
degrees of freedom. An alternative connection $A_A$ in
Eq.(\ref{eq:CS1}) is provided by the $SO(N)$ connection
$\omega_A{}^I{}_J$. In terms of the gauge groups, and the geometry
described by the two connections, this is a very different action
from the one based on $\Gamma^B_{AC}$: $SO(N)$ replaces $GL(3)$.
However, once again it must be remembered that the dynamical
variables are not the connection itself but the embedding functions
${\bf X}$. Classically, the two describe the same theory.

\vskip1pc First recall how the Chern-Simons action responds to a
change in the connection.  If $\omega_A{}^{IJ}  \to \omega_A{}^{IJ}
+ \delta \omega_A{}^{IJ}$, then
\begin{equation}
\delta S_{\rm CS} [\omega_A^{IJ}]= \int dV \, {\cal E}^{A}{}_{IJ}
\;\delta \omega_A{}^{IJ}\,,
\end{equation}
where
\begin{equation}
{\cal E}^A{}_{IJ} = \epsilon^{ABC}  \Omega_{BC\, IJ}\,.
\label{eq:A0}
\end{equation}
If the connection were the dynamical variable, the equilibria would
be provided by the flat connections. Which is not very interesting.

\vskip1pc In order to obtain the variation of $S_{\rm CS}$ with
respect to ${\bf X}$, we will exploit the method of auxiliary
variables \cite{auxil}. Following the adaptation of this method to
accommodate a normal connection presented in \cite{twist}), we treat
$\omega_A{}^{IJ}$ as a set of variables that are independent of
${\bf X}$. The price one pays is that one  must now introduce
Lagrange multipliers to enforce the constraints that connect
$\omega_A{}^{IJ}$ to the geometry. In this approach, the induced
metric $g_{AB}$, as well as the basis vectors $\{ {\bf e}_A , {\bf
n}^I \}$, are also treated as independent variables. Thus construct
the functional $S[ {\bf X}, {\bf e}_A , {\bf n}^I , \omega_A{}^{IJ}
,g_{AB}, {\cal E}_A{}^{IJ} , T^{AB} , \lambda^{IJ}, \lambda^A{}_I ,
{\bf f}^A ]$ given by
\begin{eqnarray}
S[ {\bf X},\dots]  &=&  S_{\rm CS}[ \omega_A{}^{IJ}]- \int dV
\,\left[{\cal E}_{IJ}^A \,(\omega^{IJ}_A - {\bf n}^I \cdot \nabla_A
{\bf n}^J) + {1\over 2} T^{AB} (g_{AB} - {\bf e}_A \cdot {\bf e}_B)
\right]\nonumber\\
&&\quad +  \int dV \left[ {1\over 2} \lambda_{IJ} ({\bf n}^I\cdot
{\bf n}^J - \delta^{IJ}) +  \lambda^A{}_I ({\bf n}^I\cdot {\bf e}_A)
+ {\bf f}^A\cdot ({\bf e}_A-
\partial_A {\bf X}) \right]\,. \label{eq:aux}
\end{eqnarray}
The variables ${\cal E}_A{}^{IJ} , T^{AB} , \lambda^{IJ},
\lambda^A{}_I$ and ${\bf f}^A$ are Lagrange multipliers. In this
formulation, ${\bf X}$ appears only in the last term -- the
constraint identifying the tangent vectors as derivatives of the
embedding functions. Variation with respect to ${\bf X}$ reproduces
Eq.(\ref{eq:delS}) for the action if the multiplier is identified
with the stress tensor.

\vskip1pc The Euler-Lagrange equations for  $\omega^{IJ}_A$
reproduce Eq.(\ref{eq:A0}); however, unlike the theory whose
dynamical variables are provided by the connection, in general
${\cal E}^A{}_{IJ} \ne 0$.

\vskip1pc In general, the Euler-Lagrange equations for the tangent
vectors ${\bf e}_A$ provide an expansion for ${\bf f}^A$ in terms of
the basis adapted to the surface $\{{\bf e}_A , {\bf n}^I \}$:
\begin{equation}
{\bf f}^A =  T^{AB} {\bf e}_B + \lambda^A{}_I {\bf n}^I  \,.
\end{equation}

\vskip1pc In particular, the conservation law (\ref{eq:conserv}) may
be decomposed into its tangential and normal parts by projection and
using the Gauss-Weingarten equations (\ref{eq:G}), (\ref{eq:W}):
\begin{eqnarray}
\nabla_A T^{AB} - \lambda^A{}_I \; K_A{}^{B\, I} &=& 0\,, \label{eq:const}\\
\nabla_A \lambda^A{}_I + T^{AB} K_{AB\, I} &=& 0 \label{eq:consn}\,.
\end{eqnarray}
The  normal projections (\ref{eq:consn}) are the unconstrained
Euler-Lagrange equations for the action $S_{\rm CS}$ with respect to
the embedding functions ${\bf X}$. The tangential projections
(\ref{eq:const}) are satisfied identically, providing the Bianchi
identities associated with the reparametrization invariance of
$S_{\rm CS}$.

\vskip1pc Next, note that $S_{\rm CS}$ does not depend on the
metric. Thus the variation with respect to the induced metric
$g_{AB}$ vanishes. As a consequence the metric stress tensor
$T^{AB}$ vanishes: $ T^{AB}  = 0$. This would spell triviality if
the action did not depend also on the extrinsic curvature. It does,
however, have unexpected consequences. Note, in particular, that the
stress is purely normal, which would be impossible if the action
depended only on $g_{AB}$ and $K_{AB}^I$ \cite{auxil}. In this
context, the vanishing tangential component also implies conformal
invariance.

\vskip1pc The normal component of ${\bf f}^A$ is provided by the
Lagrange multipliers,  $\lambda^A{}_I$. To determine
$\lambda^A{}_I$, consider the variation with respect to the normal
vectors ${\bf n}^I$:
\begin{equation}
\lambda^A{}_{IJ} \nabla_A {\bf n}^J + \nabla_A ({\cal E}_{IJ}^A {\bf
n}^J) +
 \lambda_{IJ} {\bf n}^J + \lambda^A{}_I {\bf e}_A =0\,.
\end{equation}
The tangential projections of this equation identify
$\lambda^A{}_I$:
\begin{equation} \lambda^A{}_I  =  - 2   \lambda^B{}_{IJ}
K^A{}_B{}^J  =  - 2   {\cal E}^B{}_{IJ} K^A{}_B{}^J  = -2
\epsilon_{ABC} \Omega^{BC}{}_{IJ}\, K^A{}_B{}^J\,.
\end{equation}
The normal projections plays no role in the conservation law; for
completeness, note that they identify the remaining Lagrange
multiplier $\lambda_{IJ}$,
\begin{equation}
\lambda_{IJ} = 2 \omega_{A(I}{}^K \lambda^A{}_{|K|J)}\,,
\end{equation}
as well as confirming the vanishing of the divergence of ${\cal
E}^A{}_{IJ}$,
the Bianchi identities associated with the $O(N)$ invariance of the
action. One concludes that the stress takes the form
\begin{equation}
{\bf f}^A =   - 2 K^{A}{}_{B}{}^J \; \epsilon^{BCD} \Omega_{CD\,IJ}
 {\bf n}^I\,.
\end{equation}
The Euler-Lagrange equations are given by
\begin{equation}
- 2\tilde\nabla_A\, (K^{A}{}_{B}{}^J \; \epsilon^{BCD}
\Omega_{CD\,IJ}) =0\,. \label{eq:ELn}
\end{equation}
A straightforward calculation indicates that Eqs.(\ref{eq:ELn})
coincide with Eqs.(\ref{eq:CK}). It can also be shown that the
difference between the two stress tensors is null. Remarkably, two
apparently very different stresses, one normal, the other tangential
describe the same theory.

\vskip1pc The tangential projections  of the conservation law for
${\bf f}^A$ (\ref{eq:const}) reads
\begin{equation} K^{A}{}_{B}{}^J \; \epsilon^{BCD} \Omega_{CD\,IJ}
K_{AE}^I =0 \,.
\end{equation}
It is also straightforward to show that these equations are
identically satisfied.

\vskip1pc Finally, note that there are no non-trivial interpolations
involving a mixing of intrinsic and extrinsic geometry. For if one
sets $(A_A)^C{}_B {}^I{}_J = \Gamma^C_{AB} \delta^I{}_J +
\omega_A^I{}_J \delta^C{}_B$, then $S_{\rm CS}[A_A] = S_{\rm
CS}[\Gamma] + S_{\rm CS}[\omega]$; all cross terms involving
$\Gamma^C_{AB}$ and $\omega_A^I{}_J$ vanish.

\section{\Large \sf Conclusions}

\vskip1pc Chern-Simons theories of three-dimensional embedded
surfaces possess a number of intriguing properties. It has been
shown that two apparently unconnected Chern-Simons theories -- one
associated with the induced Christoffel connection, the other with
the normal connection -- describe the same theory at a classical
level. Their differences have been quantified  in terms of a null
stress.
A remarkably simple geometrical property of hypersurfaces is
associated with the triviality of the corresponding Chern-Simons
theory: the stress is null. It would be interesting to examine
Eq.(\ref{eq:ckomega}) perturbatively in the neighborhood of any
hypersurface (its trivial solutions) to obtain some physical
intuition concerning its content. These results are also potentially
relevant to Deser, Jackiw and Templeton's original metric model. A
fuller examination of the connection is planned.


\vskip3pc \noindent{\Large \sf Acknowledgments}

\vskip1pc\vskip1pc \noindent Thanks to Riccardo Capovilla for
helpful comments. Partial support from CONACyT grant 51111 as well
as DGAPA PAPIIT grant IN119206-3 is acknowledged.

\vskip2pc
\noindent {\Large \sf Appendix: Null Potential for
hypersurfaces}

\vskip2pc The Cotton tensor forms a tangential null stress for a
hypersurface. Here it will demonstrated how to reconstruct this null
stress from a potential.

\vskip1pc Any null  ${\bf h}^A$ may be expanded with respect to the
adapted basis ${\bf h}^A= h^{AB}\,{\bf e}_B + h^A\,{\bf n}$.
Suppose that $h^{AB}$ is symmetric, or equivalently, that
$\epsilon_{ABC} h^{BC}=0$. Then
\begin{equation}
\nabla_A {\bf B}_B \cdot {\bf e}^B - \nabla_B {\bf B}_A \cdot {\bf
e}^B =0 \label{eq:simh}
\end{equation}
on the potential ${\bf B}_A$ given by Eq.(\ref{eq:null}). This
constraint further simplifies if ${\bf B}_A$ is tangential, ${\bf
B}^A = {\cal B}^{AB}\, {\bf e}_B$. Then Eq.(\ref{eq:simh}) reads
\begin{equation}
\nabla_B ({\cal B}^{AB}- g^{AB} {\cal B}) = 0\,, \label{eq:nabA}
\end{equation}
where ${\cal B}={\cal B}^A{}_A$.  Clearly both $g^{AB}$ and $K^{AB}-
g^{AB} K$ satisfy Eq.(\ref{eq:nabA}). However,  ${\bf h}^A$ itself
vanishes in both cases. The simplest non-vanishing choice originates
in the potential
\begin{equation}
{\cal B}^{AB}= {\cal R}^{AB} - {1\over 4} R g^{AB}\,.
\end{equation}
quadratic in the extrinsic curvature. Eq.(\ref{eq:nabA}) is
satisfied by this tensor because the Einstein tensor $G^{AB}={\cal
R}^{AB}- {\cal R} g^{AB}/2$ is divergence-free. With this choice, we
reproduce ${\bf h}^A =  C^{AB} \, {\bf e}_B$.
Note that there are no higher order tensors polynomial in $K_{AB}$
consistent with Eq.(\ref{eq:nabA}) in three-dimensions. This is
because $g_{AB}$, $K_{AB}- g_{AB} K$ and $G^{AB}$ are the only
conserved tensors polynomial in $K_{AB}$ on a three dimensional
surface. Any other solution will involve higher derivatives.

\end{document}